# Detecting high-frequency brain disorder signals using dynamic mode decomposition from EEG

**Jacob Kang**[a], **Jong-Hyeon Seo**[b,*]
[a]Fischell Department of Bioengineering, University of Maryland, College Park, MD, United States; jkang115@umd.edu
[b]School of Basic Sciences, Hanbat National University, Daejeon, Republic of Korea; hyeonni94@hanbat.ac.kr
*Correspondence should be addressed to J-H.S. (hyeonni94@hanbat.ac.kr)

**Abstract/Summary. Recent studies have reported clearly identifiable dynamical changes in the high-frequency range of EEG signals recorded during specific stimuli, such as visual or auditory inputs, or in cases of brain disorders like epileptic seizures. In this study, we utilized Dynamic Mode Decomposition (DMD) to extract consistent and persistent dynamical changes in the high-frequency band from the signals of neurologically relevant EEG channels. High-frequency DMD modes were employed as features, composing a feature table. Through post-processing, a random distribution test was performed, revealing that approximately 70% of the samples exhibited consistent high-frequency dynamics within the signal of a specific channel. Furthermore, classification experiments confirmed that the PCA components of the feature table that passed the test formed a consistent pattern that distinguished the alcohol-dependent group from the control group.**

## 1. Introduction

EEG sample signals, recorded over time intervals, typically reflect rhythmic and periodic brain activity in the low-frequency bands (e.g., $\delta$-, $\theta$-, $\alpha$-waves) [4]. EEG rhythm analysis is clinically utilized to assess neurological conditions such as epilepsy and changes in consciousness. Standard EEG interpretation focuses on identifying brainwave patterns and abnormal waveforms within specific frequency bands. [2, 10, 35].

A standard approach to analyzing brain electrical activity through frequency-based feature extraction involves examining the Spectral Density Function (SDF) and identifying predefined frequency bands that contribute the most to the overall signal variance. However, a limitation of this approach is that the exact frequency and bandwidth of oscillations can vary depending on cognitive demands [12]. Moreover, while the influence of brain waves on the spectral characteristics of EEG signals has been studied for over half a century, little is known about how biological rhythms affect brain dynamics as evaluated by modern EEG analysis techniques [24].

When neural responses induced by specific stimuli or brain disorders increase, neural activity in certain brain regions may change. During this process, the frequency or intensity of spiking may also vary. These changes in neural activity influence EEG signals, particularly by causing variations in frequency components over time, sometimes preventing the maintenance of a consistent frequency pattern. In other words, as nonstationarity increases, EEG signals exhibit continuously changing spectral characteristics over time rather than having fixed frequency components [19].

### 1.1. Nonstationary EEG signals

In addition to the rhythmic activity of the EEG, various complex patterns such as spikes and sharp waves are frequently observed in EEG recordings. These complex patterns play an important role in the diagnosis of epilepsy or seizure disorders [41]. Because the frequency and amplitude range of EEG signals continuously vary, these signals are inherently nonstationary [36]. In [34], EEG signals were treated as nonstationary signals composed of a single component with time-varying amplitude and phase, and a parametric modeling approach for these signals was proposed.

When constructing frequency-based features from nonstationary EEG signals, it is essential to consider the overall pattern of the high-frequency band. Since high-frequency signals exhibit continuous spectral characteristics (spectrum), it is difficult to divide them into specific regions and compare individual features, as is applied in traditional EEG analysis methods. Recent studies have suggested that distinctive dynamical patterns observed in the high-frequency domain of EEG signals are associated with certain neurological disorders, particularly alcoholism [9, 13, 15, 20, 23, 27]. These high-frequency activities reflect abnormal states of the nervous system and play a crucial role in assessing the brain's sensitivity to specific stimuli.

Based on the aforementioned studies, this research hypothesizes that nonstationary EEG signals behave as a single component, with their dynamic characteristics potentially emerging in the high-frequency band. To assess this, Dynamic Mode Decomposition (DMD) is applied to multichannel time-series data to simultaneously capture spatiotemporal information. By utilizing DMD modes corresponding to the high-frequency band across all channels as features derived from the EEG signals, this study seeks to validate the proposed hypothesis.

### 1.2. Dynamic Mode Decomposition

Dynamic Mode Decomposition (DMD) is a data-driven technique that extracts temporal dynamic patterns from data [30]. Originally developed in fluid dynamics, DMD has recently been applied to EEG analysis. Recent applications include pattern recognition in epileptic EEG [32], convolutional classification of Alzheimer's disease from photostimulation EEG using stacked DMD mode maps as network inputs [17], and prototype-based representations constructed from DMD mode magnitudes for dementia classification [16, 31]. By decomposing signals into eigenmodes and their corresponding eigenvalues, DMD enables the simultaneous analysis of both spatial and temporal patterns. This approach effectively captures time-dependent frequency variations in nonstationary signals and extracts modes that reflect their dynamic characteristics. Conventional frequency-based analysis methods, such as Fourier Transform (FT) and Wavelet Transform (WT), decompose signals into frequency components for analysis. While these methods are useful for identifying the frequency properties of a signal, they may have limitations in effectively tracking changes in frequency components over time, particularly in nonstationary signals. For instance, FT represents a signal as a sum of sinusoids, potentially losing information on time. WT enables time-frequency analysis, but it requires a trade-off between frequency resolution and time resolution. In contrast, DMD decomposes multichannel signals into frequency components, and each DMD mode is characterized by phase dynamics. This allows for the effective distinction between channels exhibiting strong activation at specific frequencies and those that do not. Therefore, DMD complements the limitations of traditional frequency-based analysis methods, making it a promising approach for analyzing the complex time-frequency properties of nonstationary signals.

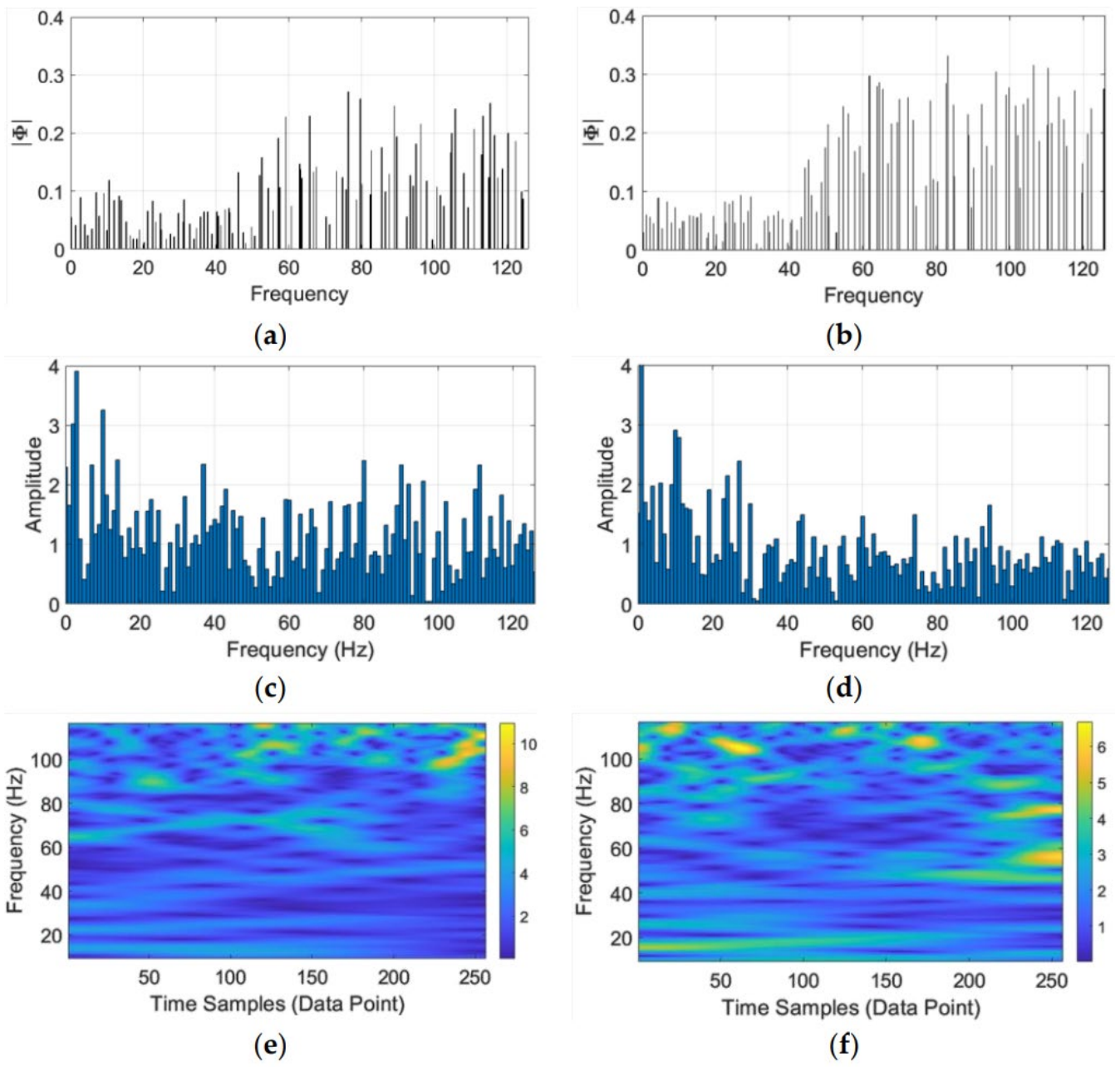

**Figure 1.** Comparison of Dynamic Mode, FFT Spectrum, and Wavelet Coefficients for Channel 55 Across Different Serial Numbers SN-1(Alcoholic), SN-236(Control).

### 1.3. Overview of Frequency- and DMD-Based EEG Analysis

In Figure 1., the high-frequency phase characteristics of Channel 55 (PO: left parieto occipital region) are compared using different analysis techniques (DMD, FFT, and WT). This region is associated with visual processing and spatial cognition and plays a crucial role in analyzing neural responses to visual stimuli. The left column ((a), (c), (e)) presents results from the Alcoholic Group, while the right column ((b), (d), (f)) shows results from the Control Group.

DMD modes extract dynamic properties based on phase information rather than signal amplitude, revealing distinct differences between the two groups, particularly in the high-frequency range (50 Hz–128 Hz). When comparing the phase distribution of DMD modes in the high-frequency range (50 Hz–128 Hz):

- In the Alcoholic Group, the phase magnitude is locally concentrated in specific frequency bands, with some range exhibiting relatively high variability.
- In the Control Group, the phase is more evenly distributed across a wider frequency range, maintaining a more uniform phase pattern rather than being restricted to specific frequency bands.

These findings suggest that in the Alcoholic Group, phase dynamics are concentrated within specific frequency bands, whereas in the Control Group, phase variations are more balanced across multiple frequency bands.

Wavelet Transform (WT), like DMD, provides an approach for analyzing the time-frequency variations of EEG signals. In Figure 1. (e) and (f), the Wavelet Coefficients of the Control Group exhibit stronger fluctuations across a broader frequency range over time, whereas the Alcoholic Group shows a pattern where activation is localized within specific frequency bands. This suggests that because Wavelet Transform can precisely track temporal changes in the signal, it serves as an additional way to visually confirm the differences in neural responses, complementing the difference of the phase information observed through DMD, and thus, the two methods reflect different aspects of neural activity. However, it is important to consider that while DMD analyzes dynamic patterns based on phase information, Wavelet Transform represents the energy distribution of the signal in a time-frequency resolution. Although Wavelet Transform provides high time-frequency resolution and captures various signal characteristics, from a feature extraction perspective, it may contain excessive information, making interpretation and application in classification models more challenging.

Figure 1. (c) and (d) present the results of Fourier Transform (FFT) Spectrum analysis in the high-frequency range. The spectrum magnitude remains relatively stable across high frequencies, with the Alcoholic Group exhibiting a relatively larger spectral magnitude than the Control Group. This pattern differs from the observations in Wavelet Transform and DMD mode analysis, highlighting a limitation

of FFT-based analysis. Since FFT converts the overall energy of a signal into frequency components, it tends to reflect differences in magnitude rather than structural complexity. As a result, even if there are structural differences in neural dynamics across specific frequency bands, FFT may overemphasize frequency components with higher signal magnitudes, regardless of underlying neural dynamics. This indicates that FFT prioritizes components with strong amplitudes rather than capturing phase characteristics or dynamic pattern variations. Consequently, if a signal has a high magnitude, the spectral magnitude may be overestimated, potentially emphasizing differences in signal intensity rather than reflecting actual differences in neural dynamics. This limitation has also been observed under matched classification pipelines, where DMD-derived representations outperformed FFT-based spectrograms when the classifier architecture, input dimensionality, and training protocol were held identical [17].

### 1.4. Ensuring the Presence of High-Frequency Neural Signals in Scalp EEG Analysis

EEG signals measured from the scalp have difficulty accurately capturing high-frequency activity due to signal attenuation and interference. However, if a signal is strong enough to be detected in scalp EEG without significant attenuation, it is likely not mere noise but rather a strong signal of neurophysiological significance. Such signals, due to the nature of high-frequency activity, tend to form distinct patterns and may contain neurophysiological features that can be meaningfully interpreted. Therefore, when analyzing high-frequency dynamics in scalp EEG signals, it is crucial to first determine whether significant signals are present in that frequency range.

For nonstationary EEG signals, DMD modes in the high-frequency band are expected to exhibit strong values following a consistent pattern in specific channels. However, directly exploring these characteristics across the entire sample is inefficient. Therefore, this study proposes a method in the post-processing stage to evaluate the distribution of DMD modes and exclude samples exhibiting a random distribution from the analysis. It should be noted that this post-processing step is distinct from the preprocessing stage, which aims to preserve as much of the original data as possible. While preprocessing refines the data for analysis using techniques such as Independent Component Analysis (ICA) [7], the post-processing approach in this study evaluates whether the DMD modes used as features are suitable for analysis, thereby selecting only reliable EEG signals.

The key distinguishing aspect of this study is that it accounts for the possibility that some samples may lack neural dynamics in the high-frequency range due to the limitations of scalp EEG. To proactively and effectively exclude such cases, validation was performed solely using DMD modes, without additional processing, and this was incorporated into the feature extraction process.

In this study, the entire set of DMD modes in the high-frequency range, which contains relatively less information compared to the low-frequency range, was utilized as features. The DMD modes of EEG signals exhibit complex patterns across different frequency bands in the low-frequency range, whereas in the high-frequency range, they form a random distribution with predominantly small values, except for certain channels containing significant information. Considering this, we propose an efficient method to exclude samples from analysis if their feature vectors from the high-frequency range follow a random distribution. By applying consistent criterion to preemptively filter out unnecessary samples, this approach enhances the reliability of analysis results and ensures more stable and consistent outcomes.

### 1.5. Overview of the Proposed Approach

In this study, Dynamic Mode Decomposition (DMD) is utilized to analyze nonstationary EEG signals. This approach enables the separation of signals by selecting an appropriate high-frequency range and effectively analyzing individual channel features and inter-channel relationships. To verify the presence of meaningful dynamics in the high-frequency range of EEG signals, DMD modes representing the high-frequency range were selected from each EEG signal and used as features. Based on these extracted features, a feature table was constructed to evaluate the differences in neural activity between the Alcoholic Group and the Control Group. In addition, this study utilizes a random test to exclude samples from the analysis if the DMD modes in the high-frequency band are determined to exhibit a random pattern. Artifacts such as electromyography (EMG) signals are characterized by appearing as asynchronous events at specific frequencies in particular channels [26]. This differs from the expectation in this study that "nonstationary EEG signals will exhibit consistent dynamical patterns across the entire high-frequency band of specific channels." Therefore, even if strong artifact signals like EMG are partially included, they will be automatically excluded from the analysis if they fail the random test, and even if included, their overall impact on the analysis remains minimal. To verify this, the analysis was performed without applying any preprocessing, and it is emphasized that this post-processing approach remains effective even when preprocessing is applied. Note that this process can be also utilized to reduce the dimensionality of feature vectors by removing channels that are identified as noise in the high-frequency range across all samples. However, to ensure an automated process that does not require additional data exploration, such additional steps were not included in this study.

## 2. Materials and Methods

This section describes the dataset, preprocessing steps, and the feature extraction and analysis methods used in this study. First, we introduce the EEG dataset used in the experiments (Section 2.1). Next, we explain the Dynamic Mode Decomposition (DMD) method, which decomposes EEG signals into dynamic components (Section 2.2). DMD modes undergo a mode selection process (Section 2.3) and are then transformed into features to construct the Feature Table (Section 2.4). To effectively summarize high-dimensional data and improve analytical efficiency, Principal Component Analysis (PCA) is applied to reduce dimensionality (Section 2.5). Finally, to account for the limitations of scalp EEG, a statistical test is performed to identify samples where high-frequency signals may not have been accurately recorded, ensuring that only reliable data is selected for final analysis (Section 2.6).

## 2.1. Dataset

Segmented EEG sample signals, recorded over short time intervals, typically reflect rhythmic and periodic brain activity. The dataset used in this study is publicly available from the UCI Machine Learning Repository (the details of the dataset [1] can be found in the link (URL: https://archive.ics.uci.edu/ml/datasets/eeg+database)) and includes EEG signals from individuals with alcohol dependence as well as from control subjects. The stimuli used in the experiment are standardized images from the Snodgrass and Vanderwart (1980) set. Participants were either shown a single stimulus (S1) with one image or two stimuli (S1 and S2) with two images presented consecutively. In the two-stimulus condition, there were matching (S2 match) and non-matching (S2 non-match) scenarios where S1 and S2 were either the same or different, respectively.

Each participant completed 120 trials, where each stimulus was presented at 1-second intervals. The data collection encompassed different paradigms including single stimulus presentation (S1), matching condition (S2 match), and non-matching condition (S2 non-match). EEG data were recorded using a 64-channel system with a sampling frequency of 256 Hz. Each sample contained 256 data points, corresponding to 1 second of data collection. The sample data provided does not include any demographic information such as age or gender. The results of classification experiments using the same dataset can be found in [29]. In this study, we used the train and test datasets redistributed via Kaggle, which were derived from the same source, to facilitate direct comparison with previously reported classification studies using the same dataset. The dataset is available at the following URL: https://www.kaggle.com/datasets/nnair25/Alcoholics. This dataset is a reprocessed version based on the original dataset provided by UCI and includes EEG signals from 16 participants (8 from the alcoholic group and 8 from the control group), excluding 4 out of the 20 original subjects. The data is provided in CSV format, with each participant contributing 60 trials of 1-second sample signals. Among the 60 trials collected from each participant, 30 were randomly split into the train set and the remaining 30 into the test set.

See Table 1 for the specified formation and allocation of the dataset. The notations used in the table refer to the recording setup and data allocation. During the data collection session, a total of 30 objects were used, divided equally into three categories: S1 (10 objects), S2 match (10 objects), and S2 non-match (10 objects). Randomly selected objects were presented to participants, and their EEG responses were recorded during each trial, resulting in 30 trials per participant.

Table 1. Formation and allocation of EEG dataset for experiment. 'SN' denotes the rearranged sample number, while 'ON' refers to the object number, where 0, 1, and 2 correspond to 'S1', 'S2 match,' and 'S2 non-match,' respectively.

| Set | Label | SN | #Participant (# averaged trial) | ON (#samples) |
|---|---|---|---|---|
| Train | alcoholic | 1 ~ 235 | 8 (30) | 0(80), 1(80), 2(75) |
| | control | 236 ~ 468 | 8 (30) | 0(80), 1(79), 2(74) |
| Test | alcoholic | 1 ~ 240 | 8 (30) | 0(80), 1(80), 2(80) |
| | control | 241 ~ 480 | 8 (30) | 0(80), 1(80), 2(80) |

## 2.2. Dynamic Mode Decomposition

One of the important usages of Dynamic Mode Decomposition (DMD) is to analyze the temporal evolution of the multidimensional signal. DMD decomposes the changing patterns of the signal into fundamental elements known as 'dynamic modes.' These modes represent the characteristics of the signal as it varies over time, and each mode represents the movement of a specific frequency within the signal [30].

The EEG signal is decomposed into a sum of signals in the DMD mode using DMD, and the filtering is performed by reconstructing the signal with only the modes that meet the filtering parameters. We briefly explain the DMD and its components, which are the results needed for decomposing the EEG signal. In addition, we specify the eigenfrequency of the decomposed mode signals to be used for filtering.

### 2.2.1. Mathematical Formulation

Dynamic Mode Decomposition (DMD) is a technique used to slice states distinguished by dynamic modes. These modes consist of empirically derived vectors, extracted directly from the data, a process elaborated in [38]. Fundamentally, DMD operates as a method for order reduction, proficient in distilling the intrinsic dynamics present in multidimensional complex systems by isolating specific frequencies, as explored in [6].

Consider a time series data

$$X \coloneqq \{\mathbf{x}(t_k)\}_{k=1}^{n} \quad (1)$$

where $\mathbf{x}(t_k)$ belongs to $\mathbb{R}^m$, and the time interval between sample points $t_{k+1} - t_k$ is fixed at $\Delta t$. For a given signal $X$ in (1), the $(ms) \times (n_s)$ shift-stack Hankel matrix $\mathbf{Y}^{(s)}$ is constructed as:

$$\mathbf{Y}^{(s)} \coloneqq \begin{bmatrix} \mathbf{x}(t_1) & \mathbf{x}(t_2) & \cdots & \mathbf{x}(t_{n_s}) \\ \mathbf{x}(t_2) & \mathbf{x}(t_3) & \cdots & \mathbf{x}(t_{n_s+1}) \\ \vdots & \vdots & \ddots & \vdots \\ \mathbf{x}(t_s) & \mathbf{x}(t_{s+1}) & \cdots & \mathbf{x}(t_n) \end{bmatrix} = \begin{bmatrix} \mathbf{y}_1 & \mathbf{y}_2 & \cdots & \mathbf{y}_{n_s} \end{bmatrix} \quad (2)$$

where $n_s \coloneqq n - s + 1$ and $s$ denotes the predetermined stack size. To encapsulate the maximal spectrum and temporal complexity of the original signal, it is imperative to maximize the dimensions of $(ms) \times (n_s)$.

The DMD algorithm accomplishes a low-rank eigendecomposition of the matrix $\mathbf{A}$ by optimally approximating $\mathbf{y}_k$ in the least squares sense, minimizing the following:

$$\|\mathbf{y}_{k+1} - \mathbf{A}\mathbf{y}_k\|. \tag{3}$$

To diminish the error, the $n_s$ column vectors are assembled into two data matrices with size $(ms) \times (n_s - 1)$:

$$\mathbf{Y}_1 = [\mathbf{y}_1 \quad \mathbf{y}_2 \quad \cdots \quad \mathbf{y}_{n_s-1}],\ \mathbf{Y}_2 = [\mathbf{y}_2 \quad \mathbf{y}_3 \quad \cdots \quad \mathbf{y}_{n_s}]$$

Subsequently, the local linear approximation can be articulated as:

$$\mathbf{Y}_2 \approx \mathbf{A}\mathbf{Y}_1 \tag{4}$$

The resolution to (4) entails discovering $\mathbf{A}$ that minimizes:

$$\|\mathbf{Y}_2 - \mathbf{A}\mathbf{Y}_1\|_{\text{Frobenius}}$$

#### 2.2.2. Eigen Decomposition and Mode Calculation

Rather than conducting the eigendecomposition of $\mathbf{A}$ directly, the DMD algorithm employs a low-dimensional surrogate, $\tilde{\mathbf{A}}$, via Singular Value Decomposition (SVD) [6, 8] of $\mathbf{Y}_1$:

$$\tilde{\mathbf{A}} = \mathbf{\Phi}\mathbf{\Lambda}\mathbf{\Phi}^{-1} \tag{5}$$

where $\mathbf{\Lambda} = \mathrm{diag}(\lambda_1, \lambda_2, \ldots, \lambda_r) \in \mathbb{C}^{r\times r}$ is a diagonal matrix containing $r\ (\leq n)$ eigenvalues of $\mathbf{A}$, and $\mathbf{\Phi} \in \mathbb{C}^{(ms)\times(r)}$ denotes the DMD modes.

In (5), each snapshot, $\mathbf{y}_{k+1}$ can be approximated as:

$$y_{k+1} \approx \tilde{\mathbf{A}} y_k$$

for $k = 1,2, \ldots, n-1$. Hence, the matrix $\tilde{\mathbf{A}}$ furnishes an approximation of the sample data, decomposing it into a unified space-time matrix:

$$\mathbf{y}_k \approx \tilde{\mathbf{A}}^{k-1}\mathbf{y}_1 = \mathbf{\Phi}\mathbf{\Lambda}^{k-1}\mathbf{c} \tag{6}$$

for $k = 1,2, \cdots n$ where $\mathbf{c}$ is a sequence of weights for which $\mathbf{y}_1 = \mathbf{\Phi}\mathbf{c}$.

Employing the components $\mathbf{\Phi}$, $\mathbf{\Lambda}$, and $\mathbf{c}$ from (5) we define a vector function $\tilde{\mathbf{x}}(t_k) \coloneqq [\tilde{x}_1(t_k), \ldots, \tilde{x}_m(t_k)]^T$ for approximating the term $\mathbf{x}(t_k)$ given by

$$\tilde{x}_i(t_k) \coloneqq \sum_{j=1}^{r} \lambda_j^{k-1} \mathbf{\Phi}_{(i,j)} c_j,\ \ k = 1,2, \ldots, n \tag{7}$$

for $i = 1,2, \ldots m$ where $\lambda_i \coloneqq \mathbf{\Lambda}_{(i,i)}$ and $\mathbf{c} \coloneqq [c_1, \ldots, c_r]^T$. An approximation $\tilde{X}$ of the original $m$-dimensional time series $X$ in (1) is provided by (7) as follows:

$$\tilde{X} \coloneqq \{\tilde{\mathbf{x}}(t_k)\}_{k=1}^{n}. \tag{8}$$

Each $\tilde{\mathbf{x}}(t_k)$ represents the data point at time $t_k$ reconstructed via DMD, minimizing the influence of noise and encapsulate the quintessential characteristics of the underlying dynamics. For additional details on the computational process, refer to [32].

#### 2.2.3. DMD Components and Eigen-frequencies

The signal $\mathbf{x}(t_k)$ is approximated to $\tilde{\mathbf{x}}(t_k)$ given in (7) by applying the DMD algorithm to the following three components: the mode matrix $\mathbf{\Phi} \in \mathbb{C}^{(ms)\times(r)}$, the eigenvalue diagonal matrix $\mathbf{\Lambda} \in \mathbb{C}^{r\times r}$, and the initial amplitude vector $\mathbf{c} \in \mathbb{C}^r$. Here, $\mathbf{\Phi}$ represents the dominant spatial structure, $\mathbf{\Lambda}^{k-1}$ represents the temporal evolution, and $\mathbf{c}$ represents the amplitude of the modes. For convenience, these three components used in the signal approximation are collectively referred to as the 'DMD components.' The discrete function $\mathbf{x}(t_k) \coloneqq [x_1(t_k), \ldots, x_m(t_k)]^T$ in (1), which defines the time series $X$, is extended to a continuous function $\mathbf{x}(t)$ using the DMD components, which is approximated by

$$x_i(t) \approx \sum_{j=1}^{r} e^{\frac{(\log\lambda_j)t}{\Delta t}} \mathbf{\Phi}_{(i,j)} c_j,\ \ t_1 \leq t \leq t_n$$

for $i = 1,2, \ldots, m$. Then the $j^{\text{th}}$ 'eigenfrequency', denoted by $\omega_j$, is given by

$$\omega_j \coloneqq \frac{\mathrm{Im}(\log\lambda_j)}{2\pi\Delta t},\ \ j = 1,2, \ldots, r \tag{9}$$

where $\omega_j$ represents the frequency, expressed in cycles per second, of the $j^{\text{th}}$ mode signal $\mathbf{\Phi}_{(:,j)} e^{\omega_j t} c_j$ corresponding to $\lambda_j$, and 'Im$(\cdot)$' denotes the imaginary part of a complex number [32].

### 2.3. DMD Mode Feature Selection Algorithm

For each EEG sample signal, denoted by $X$ in (1), we perform Dynamic Mode Decomposition (DMD) and construct the feature vector from the resulting DMD modes.

#### 2.3.1. Frequency-Based Mode Selection

The mode matrix $\mathbf{\Phi}$ in (5) and the corresponding frequencies $\omega_i$ are sorted in ascending order from the low-frequency to the high-frequency range. Here, $\mathbf{\Phi} \in \mathbb{C}^{(ms)\times(r)}$ represents the original DMD mode matrix, where $m$ is the number of channels, $s$ is the stack number, and $r$ is the total number of modes. The frequency $\omega_i$ denotes the $i$-th eigenfrequency as defined in (9). The sequence of the sorted eigenfrequency index set $\Omega$ is defined by

$$\Omega = \{i_k \in \mathbb{N}_r | k = 1, \ldots, r, \text{ and } \omega_{i_1} \leq \omega_{i_2} \leq \cdots \leq \omega_{i_r}\} \tag{10}$$

where $\mathbb{N}_r \coloneqq \{1,2, \ldots, r\}$. Then we define a frequency band FB for filtering $\Omega$ given by

$$\text{FB} \coloneqq [\omega_{\text{low}}, \omega_{\text{high}}], \tag{11}$$

where $\omega_{\text{low}}$ and $\omega_{\text{high}}$ are parameters required for frequency filtering, satisfying the condition $0 \leq \omega_{\text{low}} < \omega_{\text{high}}$. The subsequence of $\Omega$, denoted as $\mathfrak{B}_{\text{FB}}(\Omega)$, consisting of indices selected based on the given frequency filtering parameters $\omega_{\text{low}}$ and $\omega_{\text{high}}$, is defined as follows:

$$\mathfrak{B}_{\text{FB}}(\Omega) \coloneqq \{i_k \in \Omega | k = l, l+1, \ldots, u\}$$

where $l \coloneqq \mathrm{argmin}_j \left\{i_j \in \Omega \mid \omega_{\text{low}} \leq \omega_{i_j}\right\}$ and $u \coloneqq \mathrm{argmax}_j \left\{i_j \in \Omega \mid \omega_{\text{high}} \geq \omega_{i_j}\right\}$.

The high-frequency DMD modes are represented as complex values and they are provided as conjugate pairs for each eigenfrequency in a single channel. Therefore, to ensure that the modulus of each eigenfrequency does not overlap with that of any other eigenfrequency, the sequence of the indices to be finally required is constructed as follows:

$$\overline{\mathfrak{B}}_{\text{FB}}(\Omega) \coloneqq \{i_l, i_{l+2} \ldots, i_{l+v-2}, i_{l+v}\} \subset \mathfrak{B}_{\text{FB}}(\Omega), \tag{12}$$

where $v$ is an arbitrarily chosen integer satisfying $1 \leq v \leq |\mathfrak{B}_{\text{FB}}(\Omega)|$, and $\overline{\mathfrak{B}}_{\text{FB}}(\Omega)$ contains a total of round$(v/2)$ mode indices corresponding to frequencies in ascending order of

their magnitude. (Since the number of DMD modes selected through filtering varies across EEG samples, the value of $v$ must be arbitrarily chosen to standardize the dimensions of the feature vector.)

$\overline{\mathfrak{B}}_{\text{FB}}(\Omega)$ in (12) is regarded as the sequence of the filtered eigenfrequency index set, which is used to construct the final mode matrix $\mathbf{\Phi}_{\text{filtered}} \in \mathbb{C}^{(m)\times(\text{round}(v/2))}$ with filtering applied for feature extraction:

$$\mathbf{\Phi}_{\text{filtered}}(\Omega)_{(i,j)} \coloneqq \mathbf{\Phi}_{(i,\overline{\mathfrak{B}}_{\text{FB}}(\Omega)[j])} \tag{13}$$

for $i = 1,2,\dots,m$ and $j = 1,2,\dots,\text{round}(v/2)$ where $\overline{\mathfrak{B}}_{\text{FB}}(\Omega)[j]$ is the $j$-th term of the sequence $\overline{\mathfrak{B}}_{\text{FB}}(\Omega)$ and $m$ is the number of channels in the sample EEG signal $X$ in (1). $\mathbf{\Phi}_{\text{filtered}}(\Omega)$ selects only the DMD modes corresponding to rows 1 through $m$ of $\mathbf{\Phi}$, which contain the dynamics information for the original EEG signal $X$. This contrasts with selecting DMD modes for all rows of $\mathbf{\Phi}$, which is generated from the augmented series $\mathbf{Y}^{(s)}$ in (2).

#### 2.3.2. Feature Vector Generation

The feature vector is constructed by calculating the magnitude (modulus) of each mode. The modulus matrix for the filtered mode matrix $\mathbf{\Phi}_{\text{filtered}}(\Omega)$ in (12) is given as follows:

$$\left|\mathbf{\Phi}_{\text{filtered}}(\Omega)_{(i,j)}\right| = \sqrt{\left(\text{Re}\left(\mathbf{\Phi}_{(i,\overline{\mathfrak{B}}_{\text{FB}}(\Omega)[j])}\right)\right)^2 + \left(\text{Im}\left(\mathbf{\Phi}_{(i,\overline{\mathfrak{B}}_{\text{FB}}(\Omega)[j])}\right)\right)^2}.$$

for $i = 1,2,\dots,m$ and $j = 1,2,\dots,\text{round}(v/2)$. Restricting the feature construction to mode magnitude, rather than retaining phase, follows the same rationale adopted in subsequent DMD-based EEG frameworks, in which rectified magnitude descriptors were used to emphasize the spatial participation strength of each dynamic component and to reduce sensitivity to inter-subject phase variability [31]. Finally, the values of the modulus matrices from all channels are concatenated to form a long feature vector $\mathbf{f}$, as follows:

$$\mathbf{f} = \text{vec}(|\mathbf{\Phi}_{\text{filtered}}(\Omega)|)^T, \tag{14}$$

where $\text{vec}(\cdot)$ denotes an operation that concatenates all columns of a given matrix into a single column vector.

### 2.4. Feature Table Construction

We construct a Feature Table from both the training EEG data and the test EEG data. The given EEG dataset $\{X_1, X_2, \dots, X_p\}$ consists of EEG sample signals $X_i$ with $m$ channels, $i = 1,2,\dots,p$, in the form of time series, as defined in (1). We apply the DMD Mode Feature Selection Algorithm described in Section 2.3 to each EEG sample signal $X_i$ to generate its feature vector $\mathbf{f}_i$ in (14), and finally, we define the feature table $\mathbf{F}$ as follows:

$$\mathbf{F} \coloneqq \begin{bmatrix} \mathbf{f}_1 \\ \mathbf{f}_2 \\ \vdots \\ \mathbf{f}_p \end{bmatrix}, \quad \mathbf{f}_i \in \mathbb{R}^d \tag{15}$$

where $d$ is the dimension of the feature vector given by $d \coloneqq m \times (\text{round}(v/2))$ and $v$ is an integer for which $v \leq \min\{|\mathfrak{B}_{\text{FB}}(\Omega_1)|, \dots, |\mathfrak{B}_{\text{FB}}(\Omega_p)|\}$ with the sequence of the sorted eigenfrequency indices $\overline{\mathfrak{B}}_{\text{FB}}(\Omega_i)$ defined in (12).

### 2.5. Data Normalization and Dimensionality Reduction

This study asserts that nonstationary EEG signals obtained by measuring brain activity in patients with specific irregular stimuli or brain disorders exhibit a characteristic pattern where certain connected channels in the high-frequency range are strongly activated due to brain functionality. Notably, such patterns tend to be repeatedly observed in similar types of EEG signals. Consequently, dimensionality reduction techniques can be effectively utilized to analyze these signals more efficiently.

Before applying PCA for dimensionality reduction on the Feature Table $\mathbf{F}$ used for training, each feature is normalized to ensure a consistent mean and variance. The mean vector $\boldsymbol{\mu} = (\mu_1, \dots, \mu_d)$ and the standard deviation vector $\boldsymbol{\sigma} = (\sigma_1, \dots, \sigma_d)$ of the features are computed as follows:

$$\mu_j = \frac{1}{p}\sum_{j=1}^{p} \mathbf{F}_{(:,j)}, \quad \sigma_j = \sqrt{\frac{1}{p}\sum_{j=1}^{p}\left(\mathbf{F}_{(:,j)} - \mu_j\right)^2} \tag{16}$$

where $\mathbf{F}_{(:,j)}$ represents the $j^{\text{th}}$ feature (or column vector) in the feature table $\mathbf{F}$ in (15), and $p$ is the total number of samples in the training data. Subsequently, each feature $\mathbf{F}_{(:,j)}$ in the training data is normalized as follows:

$$\tilde{\mathbf{F}}_{(:,j)} = \frac{\mathbf{F}_{(:,j)} - \mu_j}{\sigma_j}$$

where $\tilde{\mathbf{F}}_{(:,j)}$ is the $j^{\text{th}}$ normalized feature. Then we obtain the normalized feature table

$$\tilde{\mathbf{F}} \coloneqq \left[\tilde{\mathbf{F}}_{(:,1)}, \dots, \tilde{\mathbf{F}}_{(:,d)}\right]$$

For $\tilde{\mathbf{F}}$ based on $\mathbf{F}$, Principal Component Analysis (PCA) [21] is performed to generate the dimensionally reduced Feature Table $\overline{\mathbf{F}}$ as follows:

$$\overline{\mathbf{F}} = \tilde{\mathbf{F}} \cdot \mathbf{W} \tag{17}$$

where $\mathbf{W} \in \mathbb{R}^{d\times l}$ is the matrix of principal components obtained from the PCA process, and the number of components is selected to retain the desired variance or dimensionality.

### 2.6. Selection of Feature Vector

The DMD modes extracted from EEG signals are represented as complex numbers, and in this study, their modulus values are used for analysis. If the signals in the high-frequency range are insignificant random signals, such as white noise, the modulus of the DMD modes in this range is likely to follow a Rayleigh distribution [14]. Therefore, the features are assumed to follow a Rayleigh distribution, and a goodness-of-fit test is performed. Based on this test, only the feature vectors that are unlikely to follow a Rayleigh distribution and thus have a higher probability of containing meaningful information, are selected for analysis.

#### 2.6.1. Characteristics of Complex Modulus Data

The modulus of a complex number $z = a + bi$ is defined as:

$$|z| = \sqrt{a^2 + b^2},$$

where $a$ and $b$ are the real and imaginary parts, respectively. If the real and imaginary parts follow a normal distribution with mean 0 and equal variance, the modulus $|z|$ follows a Rayleigh distribution. The probability density function (PDF) of the Rayleigh distribution is given by:

$$f(r;\sigma) = \frac{r}{\sigma^2} e^{-\frac{r^2}{2\sigma^2}},\ \ r \geq 0$$

where $\sigma$ is the scale parameter, which is related to the variance of the underlying normal distributions of the real and imaginary parts.

#### 2.6.2. Rayleigh Distribution Goodness-of-Fit Test

To assess the goodness-of-fit for the Rayleigh distribution, we first estimate the scale parameter $\sigma$ of the data. Based on the estimated $\sigma$, we evaluate the fit of the data to the Rayleigh distribution. For the fitted Rayleigh distribution, we perform the Kolmogorov-Smirnov (KS) test [25].

The $p$-value of the test is defined by the following function to evaluate whether the data fits the Rayleigh distribution:

$$\mathcal{P}(\mathbf{f}_i) = \int_{-\infty}^{\infty} \left|F_{\text{empirical}}(r) - F_{\text{Rayleigh}}(r;\sigma)\right| dr,$$

where $F_{\text{empirical}}$ is the empirical cumulative distribution function (ECDF) of the data, and $F_{\text{Rayleigh}}(r;\sigma)$ is the cumulative distribution function (CDF) of the estimated Rayleigh distribution with the scale parameter $\sigma$. The $p$-value is used to assess the goodness-of-fit, with a smaller $p$-value indicating a better fit to the Rayleigh distribution.

#### 2.6.3. Feature Filtering Criteria

Features that do not follow a Rayleigh distribution are considered likely to contain meaningful signals and thus are retained. The set of feasible indices $\mathcal{R}$ for the feature table $\mathbf{F}$ is defined as:

$$\mathcal{R}(\mathbf{F}) = \{i|\mathcal{P}(\mathbf{f}_i) > \alpha\}, \quad (18)$$

where $\alpha$ is the significance level ( $\alpha = 0.05$ ). This set includes the indices of the features that do not follow the Rayleigh distribution and are therefore considered as containing potentially significant information for further analysis.

Figure 2 illustrates the DMD-based feature extraction and selection process. After performing Dynamic Mode Decomposition (DMD) on the EEG signals, High-Frequency Band-Pass Filtering is applied to select meaningful DMD modes. Next, Principal Component Analysis (PCA) is used to reduce dimensionality, followed by normalization (mean and standard deviation) to construct the feature vectors. Finally, a Feasible Test is conducted to select reliable samples, leading to the creation of the final Feasible Feature Table for training and testing.

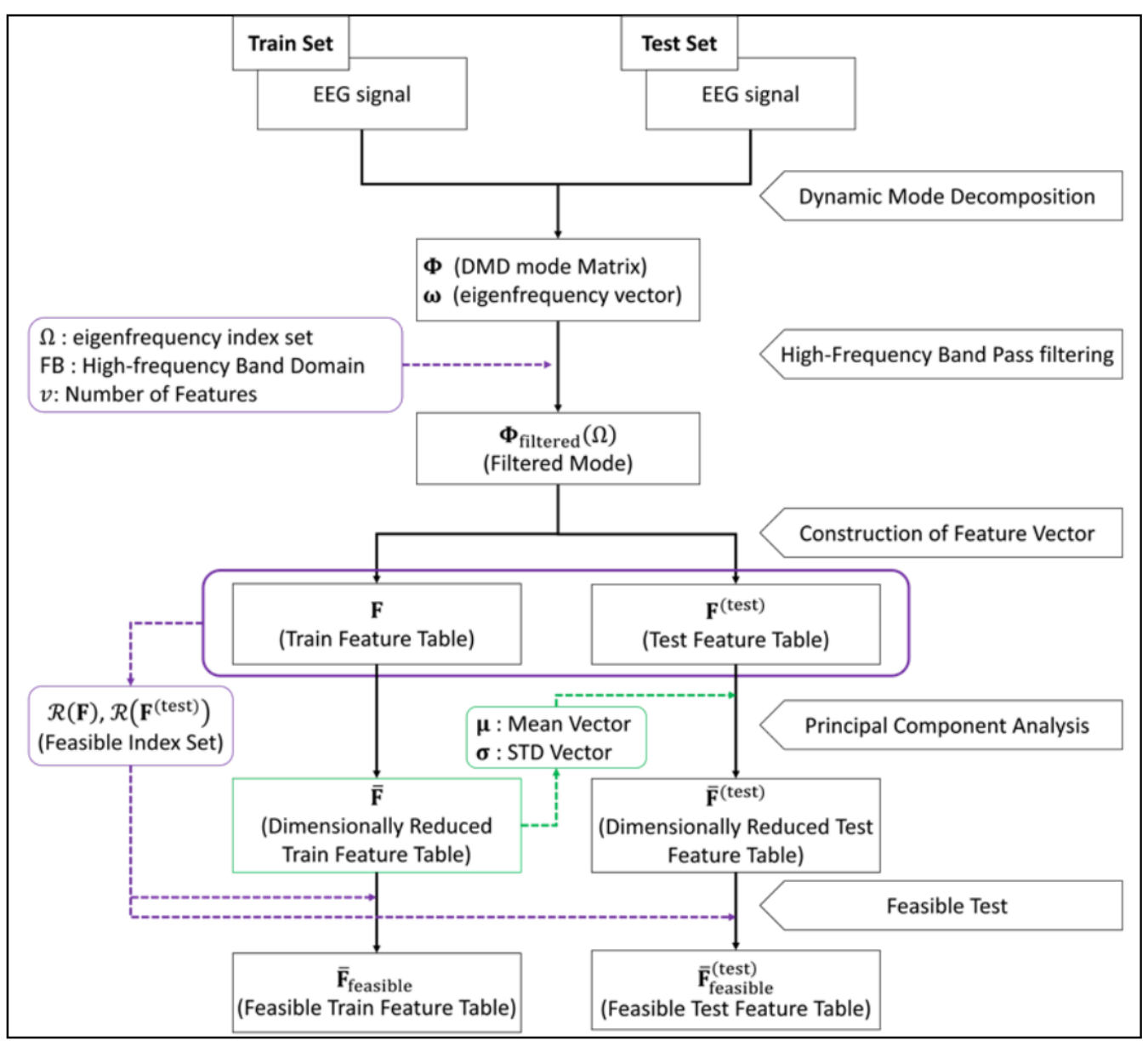

**Figure 2.** Schematic Diagram of Feature Extraction and Selection using DMD and PCA

Figure 3 provides a visualization example of features extracted from EEG sample signals classified as feasible and infeasible. Figure 3 (a) and (d) show the modulus of DMD modes for EEG sample signals from Serial Numbers 1 (SN-001) and 100 (SN-100), respectively, with $\omega$ representing the eigenfrequencies in the DMD mode index set. Figure 3 (b) and (e) display histograms of the modulus of high-frequency DMD modes and their fits to Rayleigh distributions for SN-001 and SN-100. Figure 3 (c) and (f) present Q-Q plots comparing the empirical distribution of the modulus of DMD modes to the theoretical Rayleigh distribution, visually assessing the goodness-of-fit.

The histograms represent the modulus of features selected by the DMD Mode Selection Algorithm, while the red lines indicate the probability density function (PDF) of the fitted Rayleigh distribution. In Figure 3 (b), the feature distribution does not align with the Rayleigh distribution, indicating that $\mathbf{f}_1$ is feasible. Conversely, in Figure 3 (d), $\mathbf{f}_{100}$ is deemed infeasible and should be excluded. For the sample data labeled SN-001, it is evident that the high-frequency dynamics contain meaningful information, as the feature histogram does not conform to the Rayleigh distribution, confirming its feasibility. Conversely, for SN-100, the high-frequency signals appear to be randomly distributed. This may indicate either the imprecision of the EEG measurement device or the lack of active brain responses to the stimulus. The feature histogram in this case fits the Rayleigh distribution, indicating infeasibility. Additionally, the modulus of the DMD modes reveals channels with significant information in specific segments of the low-frequency range.

In the Q-Q plot of Figure 3 (c), the empirical quantiles deviate significantly from the theoretical quantiles (Rayleigh distribution) in the upper quantile region. This indicates that the high-frequency components of SN-001 do not follow a simple Rayleigh distribution and may contain certain neurological signals. In contrast, the Q-Q plot of Figure 3 (f) shows that the empirical quantiles are more aligned with the theoretical quantiles and tend to stay close to a straight line. This pattern reveals that the high-frequency components of SN-100 exhibit statistical properties like a Rayleigh distribution, implying a higher likelihood of random noise-like signals. In summary, SN-001 is likely to contain highly structured, neurologically meaningful abnormal signals, whereas SN-100 may represent a signal with relatively stronger random components. This observation suggests that for infeasible EEG signals, conventional analysis methods may be more effective in extracting meaningful insights.

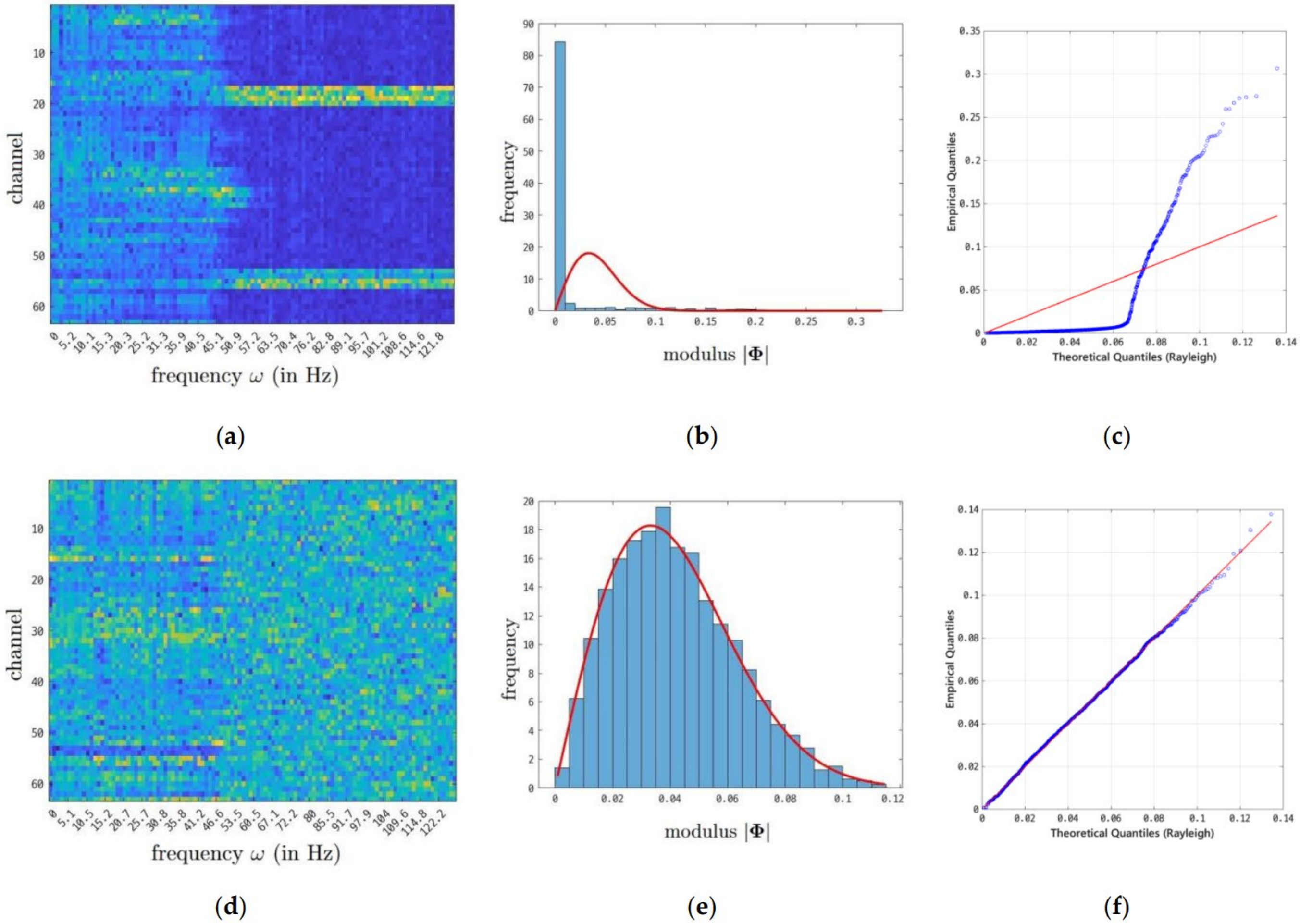

**Figure 3.** Significant and Non-Significant DMD Modes in the High-Frequency Range of EEG Data. (a) and (d) the modulus of DMD modes. (b) and (e): the histograms of the features and their fits to Rayleigh Distributions. (c) and (f): Q-Q Plot for assessing Rayleigh Distribution Fit of the features.

# 3. Experiment

In [29], classification experiments were conducted on alcohol-dependent patients and the general population using the same dataset as in this study. The goal of this study is to verify whether distinctive dynamics in the high-frequency range can effectively distinguish between the two classes. Therefore, only high-frequency components from the EEG signals in the dataset were featured for the experiment. To achieve this, DMD modes were used to extract features from the frequency band above 50 Hz, and classification experiments were performed based on these features. Additionally, the distribution characteristics of the data were examined through the visualization of the extracted high-frequency features, and the differences between the classes were analyzed. The parameters configured for the featuring with the given data are summarized in Table 2.

Table 2. List of parameters and their descriptions.

| Symbol | Value | Description |
|---|---|---|
| m | 64 | Number of channels in an EEG sample signal in (2) |
| n | 256 | Number of data points in one-second EEG sample (256 Hz) in (2) |
| s | 10 | Predetermined stack size in (2) |
| r | 220 | Number of DMD modes in (5) |
| FB | [50, 120] | Frequency band for filtering in (11) |
| v | 50 | Number of features in (12) |
| l | 6 | Number of PCA components in (17) |

In this study, a Support Vector Machine (SVM) was used to train and evaluate the classification model. SVM learns the optimal decision boundary based on quadratic programming [20]. For model training, the feature-reduced feature table $\overline{\mathbf{F}}$ in (17) from the train dataset is used, and its corresponding label vector y is utilized to find the optimal decision hyperplane. Furthermore, during the actual training process, only the indices belonging to the feasible set in the feature table are considered for calculation. Evaluation is carried out using the test dataset, and the performance of the model is measured using metrics such as accuracy, precision, and recall.

A classification experiment aims to validate the assumption that a specific channel exhibits a consistent dynamical pattern across the entire high-frequency range. It is conducted by selecting feasible feature vectors that have passed a random classification test, which is used to exclude artifacts such as EMG—characterized by strong signals in certain channels and frequency ranges—from the analysis. In other words, this experiment seeks to verify whether the features extracted from the high-frequency domain display the same pattern and distribution in both the training set and the test set when only the signals validated through post-processing are analyzed. It should be emphasized that this is not a proposal for a specific classification algorithm aimed at improving accuracy.

## 3.1. Training Data

During the training process, SVM solves the following optimization problem:

$$\min_{\mathbf{w},b} \frac{1}{2}\|\mathbf{w}\|_2^2 \text{ such that } y_i\left(\mathbf{w}^T\overline{\mathbf{f}}_i + b\right) \geq 1,\ i \in \mathcal{R}(\mathbf{F}),$$

where the label vector $\mathbf{y}$ is given by

$$\mathbf{y} = \left[y_1, y_2, \dots, y_{|\mathcal{R}(\mathbf{F})|}\right],\ \ y_i \in \{0,1\},$$

and $y_i = 0$ is an Alcoholic Sample and $y_i = 1$ is a Control Sample.

## 3.2. Test Data

The test features $\mathbf{F}_{(:,j)}^{(\text{test})}$, where $j = 1,2,\dots,d$, are normalized using the mean vector $\boldsymbol{\mu}$ and standard deviation vector $\boldsymbol{\sigma}$ obtained from the training feature table $\mathbf{F}$ based on the calculation in (16), as follows:

$$\tilde{\mathbf{F}}_{(:,j)}^{(\text{test})} = \frac{\mathbf{F}_{(:,j)}^{(\text{test})} - \mu_j}{\sigma_j} \tag{19}$$

The normalized test feature table $\tilde{\mathbf{F}}^{(\text{test})}$ and the matrix of principal components $\mathbf{W}$ derived from $\mathbf{F}$ in (18) are input into (19) to compute the dimension-reduced feature table $\overline{\mathbf{F}}^{(\text{test})}$. This reduced feature table is subsequently used for prediction:

$$\hat{y}_i^{(\text{test})} = \operatorname{sgn}\left(\mathbf{w}^T\overline{\mathbf{f}}_i^{(\text{test})} + b\right)$$

for $i \in \mathcal{R}\left(\mathbf{F}^{(\text{test})}\right)$, where $\mathcal{R}(\cdot)$ is feasible index set in (17) and $y_i^{(\text{test})} \in \{0,1\}$ is the predicted class label for $\overline{\mathbf{f}}_i^{(\text{test})}$ ($i$-th feature vector of $\overline{\mathbf{F}}^{(\text{test})}$).

In Figure 4, the distribution of dimensionally reduced features shows patterns that make it difficult to achieve a linear separation between the classes. Therefore, in this study, the kernel method in [20] of the Support Vector Machine (SVM) was applied to classify the alcoholic group and the control group. The kernel used was the RBF (Radial Basis Function) kernel, a nonlinear kernel provided by default in MATLAB. The experiment was conducted using MATLAB 2022(b), particularly employing the 'fitcsvm' function for SVM classification tasks.

## 3.3. Criteria for Parameter Selection

The parameters related to feature extraction and selection for the given data are summarized in Table 2. The criteria for selecting the frequency band for filtering (FB in (11)) and the number of selected DMD modes ($v$ in (12)), as presented in Table 2, are as follows:

Frequency Band for Filtering (FB): Figure 3 (a) and (d) show the frequency distribution of the DMD modes. In some channels, neural activity is maintained even above 50 Hz, while in others, only random noise patterns are observed. These patterns are consistently observed across the entire Train/Test dataset. Therefore, in this study, DMD modes above 50 Hz are selected and considered as neurologically significant feature patterns in the high-frequency range. The influence of the retained frequency range on DMD-based classification has been examined directly in a related setting, where excluding the delta band improved

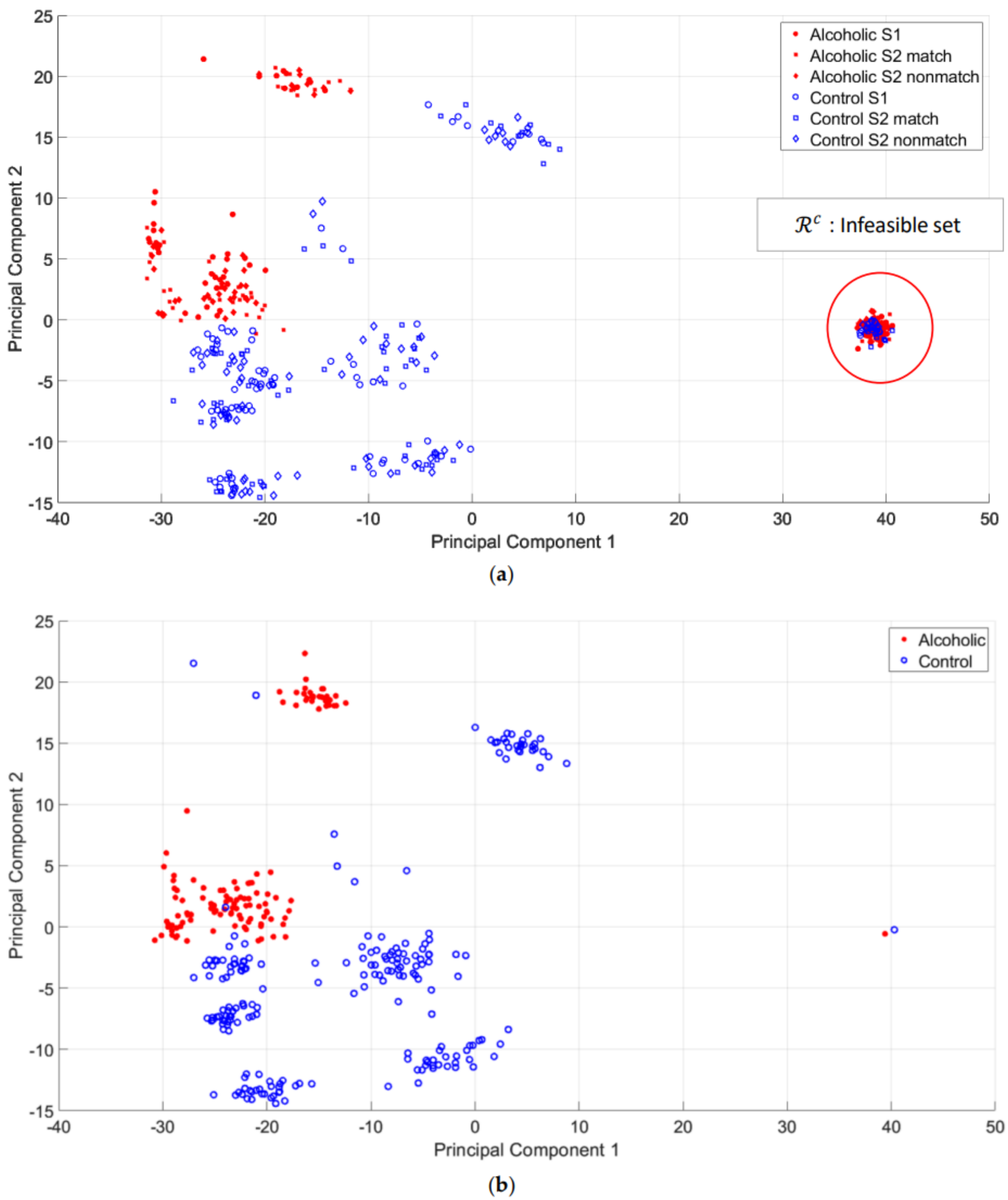

**Figure 4.** Dimensionality-Reduction of Feasible Feature Tables (a) $\mathbf{F}$ in (17) and (b) $\mathbf{F}^{(\text{test})}_{\text{feasible}}$ in (20), generated by Principal Component Analysis

Alzheimer's disease classification from eyes-open EEG under an otherwise identical pipeline [17], indicating that the choice of filtering band is a substantive analytical decision rather than a purely technical one.

Number of Selected DMD Modes ( $v$ ): To select an appropriate value for $v$, the total number of DMD modes with intrinsic frequencies above 50 Hz was calculated for each EEG signal, and the average was obtained. As a result, among the total 220 modes ($r = 220$), approximately 100 modes were found on average. Based on this, half of the modes ( $v = 50$ ) were selected as features. Due to the nonstationarity of EEG signals, the selection of the filtering band has a relatively minor impact on the results in the high-frequency range analysis. That is, even if the value of

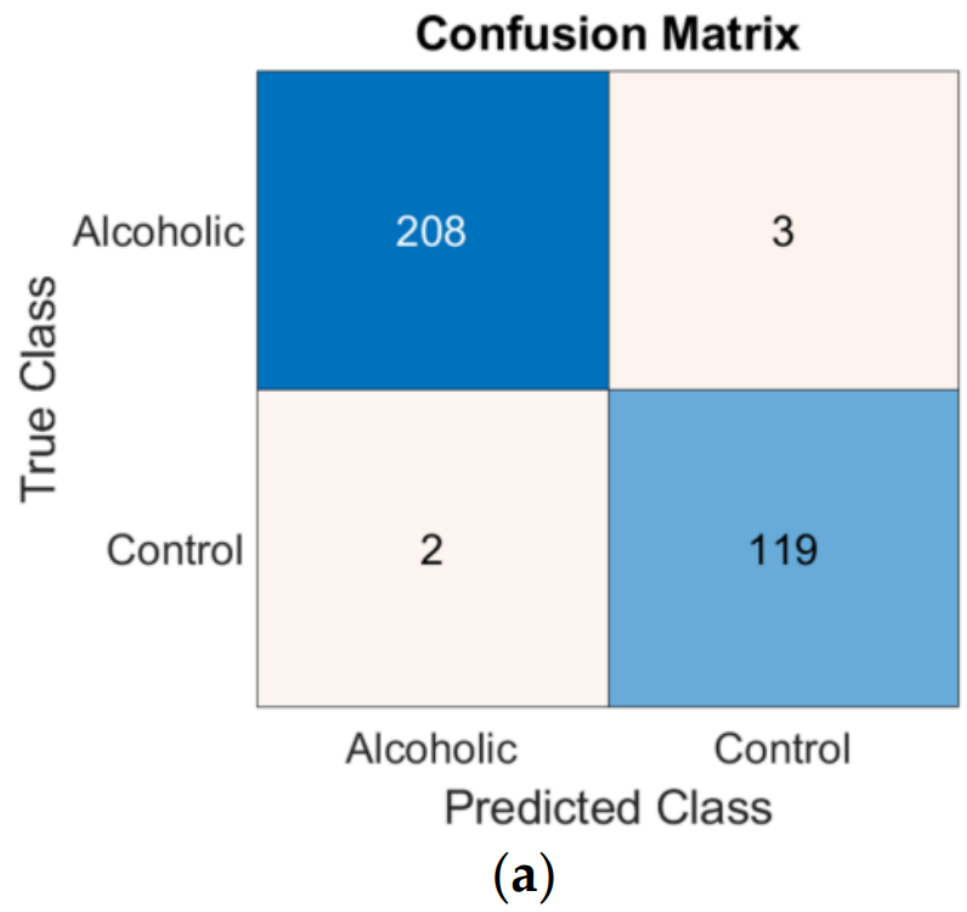

(a)

| Metric | Value (%) |
|---|---|
| Sensitivity (Recall) | 98.6 |
| Specificity | 98.6 |
| Precision | 99.0 |
| Accuracy | 98.5 |

(b)

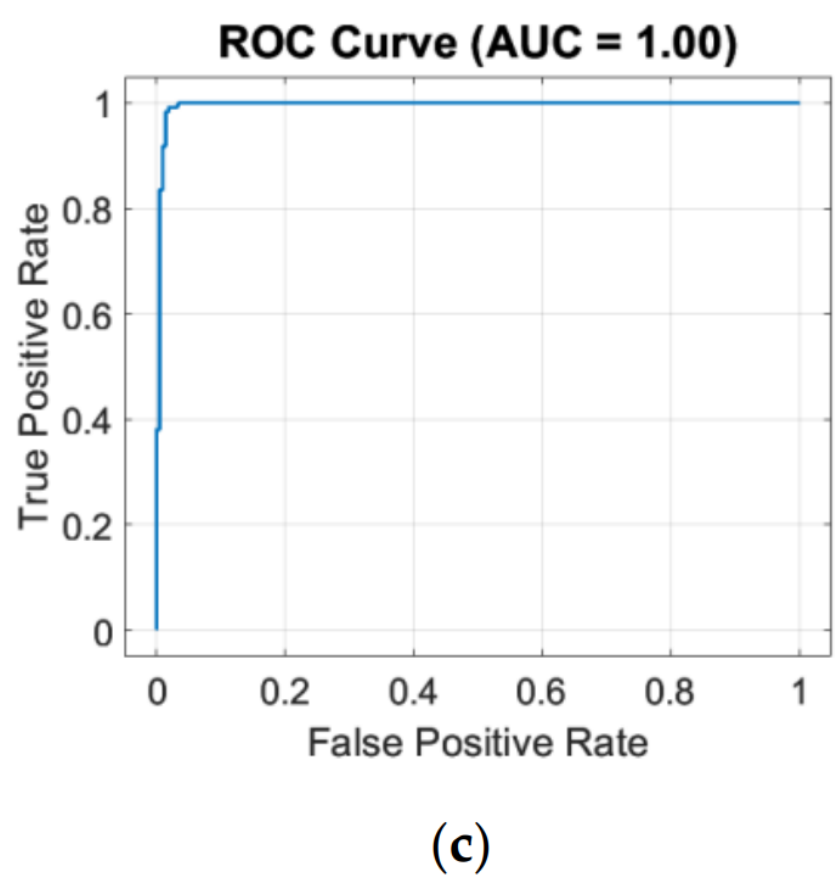

(c)

**Figure 5.** Confusion Matrix (a), Key Performance Metrics (b) and ROC curve (c) for $\overline{\mathbf{F}}^{(\text{test})}_{\text{feasible}}$

$v$ is slightly larger or smaller than 50, there is no significant difference in the reliability and accuracy of the overall analysis.

## 4. Result

Figure 4 visualizes the distributions of the dimensionally reduced feature table $\overline{\mathbf{F}}$ in (17), constructed from the train set, and the dimensionally reduced feasible feature table defined as

$$\overline{\mathbf{F}}^{(\text{test})}_{\text{feasible}} := \overline{\mathbf{F}}^{(\text{test})}_{\left(\mathcal{R}\left(\mathbf{F}^{(\text{test})}\right),:\right)}, \tag{20}$$

constructed from $\mathbf{F}^{(\text{test})}$ in (19), in (a) and (b), respectively. Here, $A(a,:)$ denotes the partial matrix reconstructed by selecting only the rows corresponding to the index sequence $a = \{a_i\}_{i=1}^{k(\leq m)}$ from $A \in \mathbb{R}^{m\times n}$. In these figures, 2D scatter plots represent the two principal components ($l = 2$). Red points correspond to the Alcoholic group, while blue points represent the Control group, highlighting the distribution differences between the two groups. Among the 486 training samples (Alcoholic: 235, Control: 233), the feasible features include 118 samples for the Alcoholic group and 204 samples for the Control group, and among the 480 test samples (Alcoholic: 240, Control: 240), the feasible features include 211 samples for the Alcoholic group and 121 samples for the Control group. Both figures confirm that the feasible features distinctly separate the Alcoholic group from the Control group.

Figure 4 (a) presents the PCA results for the entire train feature table, where the infeasible set $\mathcal{R}^C$ forms a normally distributed cluster in the Principal Component space (with most features in $\mathcal{R}^C$ identified as infeasible through the feasibility test, except for a few outliers). In contrast, Figure 4 (b) displays the dimensionally reduced feasible feature table for the test samples, where infeasible features are excluded from the PCA transformation. Despite passing the feasibility test, some features are observed to occupy regions in the Principal Component space typically associated with infeasible features.

Thus, since some infeasible data may potentially pass the feasibility test, PCA transformation for the train feature table must be performed on the entire feature table, regardless of the feasibility test results. This ensures that features from the test data, which may pass the feasibility test despite being infeasible, are positioned in regions of the Principal Component space associated with infeasible features when dimensionally reduced through PCA, without interfering with the regions where feasible features are distributed. Consequently, PCA for the train data should be applied to the entire dataset, independent of the feasibility test. However, during the training and testing processes for the train and test feature tables, only feasible features are utilized for both.

In the principal component space for PCA in Figure 4, considering the feasible features of the Train and Test sets, the Alcoholic group is divided into two subgroups. Compared to the Control group, the Alcoholic group exhibits a relatively more concentrated distribution. Both groups show significant high-frequency responses in the same brain region associated with visual stimuli, the Occipital Lobe. However, the distribution characteristics in the PCA space differ. The Alcoholic group forms a dense cluster region with the 8 experimental participants closely grouped together. In contrast, the Control group shows individual differences, with each participant distributed across different cluster regions. These results suggest that the EEG signal patterns of the Alcoholic group are distinct from those of the Control group. Specifically, the high-frequency neural dynamics of alcohol-dependent patients are likely to exhibit a more structured (clustered) pattern compared to the Control group.

Table 3. Confusion Matrix and Corresponding Performance Metrics

| | **Predicted Positive** | **Predicted Negative** | **Performance Metric** | **Trade-Off Metric** |
|---|---|---|---|---|
| Actual Positive | True Positive (TP) | False Negative (FN) | Sensitivity (Recall) TP / (TP + FN) | Miss Rate FN / (TP + FN) |
| Actual Negative | False Positive (FP) | True Negative (TN) | Specificity TN / (TN + FP) | Fall-Out 1 - Specificity |
| Performance Metric | Precision TP / (TP + FP) | Accuracy (TP + TN) / (TP + TN + FP + FN) | | |

These findings become more evident through the classification experiment. Figure 5 presents the classification results of a model trained on the train feature table ($\overline{\mathbf{F}}$ in (17)) and tested on the feasible test feature table ($\overline{\mathbf{F}}^{(\text{test})}_{\text{feasible}}$ in (19)). It also provides a quantitative evaluation of model performance based on the confusion matrix and key performance metrics, displayed in Table 3, including Precision, Recall, and Accuracy.

The overall classification accuracy on the test data was over 98%, which is highly consistent with the classification accuracy (98.99%) reported in [29]. The primary objective of this study is to verify whether high-frequency dynamics can differentiate alcoholic patients from healthy individuals. Therefore, the lower overall accuracy when considering all samples does not undermine the core intent of the research.

The test results may vary if the Train/Test dataset split criteria are modified (e.g., through cross-validation). However, since this study utilizes a pre-balanced dataset, additional cross-validation experiments were not conducted. Moreover, the improvement in model training accuracy and the generalizability of the results are more significantly influenced by the ratio of feasible and infeasible samples within the training set and the quality of feasible samples rather than by a simple Train/Test data split. Hence, conducting additional experiments would require a more rigorous and complex dataset configuration, which was beyond the scope of this study.

The primary focus of this study is not to evaluate the performance of the classification model but to verify whether distinctive dynamics exist in the high-frequency EEG signals of alcohol-dependent patients that differentiate them from the general population. The experiments conducted in this study provide sufficient evidence to support this hypothesis. As shown in Figure 2 and Figure 3, in samples where high-frequency dynamics are present, scalp EEG also exhibits clear neurophysiological differences. Therefore, the high classification performance of the model is valid and suggests that DMD-based high-frequency features effectively capture specific neurophysiological patterns.

In conclusion, high classification performance is observed in samples where significant high-frequency dynamics are present. This demonstrates that distinctive high-frequency patterns exist in the EEG signals of alcohol-dependent patients compared to the control group in response to specific visual stimuli.

## 5. Discussion and Conclusion

The experimental results demonstrated that distinctive dynamical changes in the high-frequency range of EEG sample signals can be effectively identified using Dynamic Mode Decomposition (DMD). Notably, when the modes extracted from the high-frequency range via DMD were used as features to construct a Feature Table and applied to classification experiments with a Support Vector Machine (SVM), the EEG signals of alcoholic patients and the control group were distinguished with an accuracy exceeding 98%. These findings suggest that analyzing the dynamical characteristics in the high-frequency range offers significant potential for EEG-based classification and diagnostic applications.

In all EEG signals included in the dataset, the low-frequency range tends to exhibit relatively distinct mode magnitudes in the occipital region channels (Channel-55 and Channel-56) compared to other channels, as shown in Figure 3. In the feasibility test, DMD-based analysis confirmed that in both the Alcoholic Group and Control Group, more than 70% of the data exhibited activation in the high-frequency range (above 50 Hz) in occipital region channels in response to visual stimuli. Additionally, in the Alcoholic Group, strong high-frequency dynamics were observed in Channels 17 (C3) and 18 (C4), which did not appear in the Control Group. C3 and C4 correspond to electrode positions in the central region of the left and right hemispheres, respectively, and are closely associated with the primary motor cortex (M1). This indicates that in the Alcoholic Group, additional high-frequency dynamics unrelated to visual stimuli were present.

Conventional EEG analysis techniques, such as Fourier Transform and Wavelet Transform, analyze all frequency bands uniformly, making it challenging to effectively isolate meaningful signals in specific high-frequency ranges. In contrast, DMD-based analysis separates frequency-specific dynamics, allowing for the quantitative extraction of distinct neurophysiological patterns, making it a more advantageous approach. Thus, the DMD-based high-frequency features used in this study effectively capture the distinctive patterns observed in the neural responses of the Alcoholic Group, which can be interpreted as a key factor in achieving the high classification performance of 98%.

These results are not merely due to differences in data distribution but may also be linked to strong neural firing responses to visual stimuli in the occipital region, as well as the additional high-frequency patterns observed in the Alcoholic Group. In the Alcoholic Group, strong high-frequency dynamics were observed not only in the occipital region but also in the central region, including C3 and C4. As reported in [8], in patients with schizophrenia, gamma amplitude in EEG signals significantly increases during positive symptoms such as hallucinations. This suggests that such patterns may also be related to sensory disturbances and visual hallucinations in alcohol-dependent individuals. Therefore, high-frequency responses observed across multiple brain regions, including the occipital lobe, may serve as a key feature distinguishing the Alcoholic Group from the Control Group. The findings of this study emphasize that high-frequency components in EEG can provide valuable information for detecting neurological abnormalities. Additionally, they demonstrate that DMD-based analysis is a more effective tool than conventional methods for identifying distinctive neural dynamics in the high-frequency range.

However, the accuracy of high-frequency measurements in EEG signals can vary depending on the performance of the recording equipment, the intensity of the stimuli, and the state of brain activity. High-frequency signals are particularly sensitive to external factors, which may influence experimental outcomes. In this study, a feasible region was defined to account for these potential limitations, enabling the extraction of reliable characteristics from the high-frequency range. Nonetheless, achieving consistently high accuracy across diverse EEG recording conditions remains challenging. Therefore, further validation under various measurement conditions and experimental environments is necessary.

To identify neurologically meaningful high-frequency signals and remove high-frequency artifacts such as electromyography (EMG), traditional approaches primarily relied on Independent Component Analysis (ICA) [7]. ICA is primarily used for signal separation; however, its effectiveness depends on proper preprocessing settings. In particular, high-pass filtering parameters require careful adjustment, as they may influence the preservation of neurologically significant signal components [26]. Additionally, because high-frequency signals overlap with the frequency bands of muscle activity, there is a concern that muscle-induced artifacts could be mistaken for high-frequency neural activity, making it crucial to remove these artifacts through post-processing [40]. Therefore, this study proposes an additional post-processing approach. Our proposed post-processing analyzes the distribution of DMD modes directly without any supplementary analysis techniques. It automatically excludes cases from the analysis where the high-frequency signals have been contaminated during preprocessing or when the expected signal pattern in the high-frequency band is absent due to inaccuracies in the EEG measuring device. A caveat of this approach is that samples excluded as randomly distributed are not necessarily uninformative. Recent work suggests that apparently weak or fragmented EEG configurations may still contain recurrent, class-associated structure when encoded relative to learned prototype patterns rather than evaluated by distributional fit alone [16]. The feasibility test adopted here should therefore be understood as a conservative reliability filter rather than a definitive separation of signal from noise.

Dynamic Mode Decomposition (DMD) is a useful technique for analyzing the dynamic patterns of data to identify the characteristics of high-frequency signals. By exploring data through DMD, both the temporal variations and spatial distribution of signals can be considered simultaneously, providing intuitive insights into the underlying neural dynamics. For example, in this study, DMD-based data exploration revealed that high-frequency signals primarily appear in the occipital lobe, which is responsible for visual processing. Additionally, in alcohol-dependent patients, extra activation was observed in specific channels such as C3 and C4. This allowed for a preliminary verification that high-frequency artifacts had minimal influence on the overall dataset.

Due to the limitations of scalp EEG, the accurate recording of intrinsic high-frequency dynamics cannot be guaranteed. Thus, while the proposed methodology can be effectively applied to many types of EEG signals, it has a limitation in that it cannot be guaranteed to work universally for every kind. To overcome this limitation, this study incorporated a process in the classification procedure to select only samples that contain meaningful signals in the high-frequency range. As a result, when analyzing only the selected signals, a high classification accuracy was achieved. This suggests that if high-frequency dynamics can be reliably measured in EEG signals, there is a high likelihood of obtaining reliable analytical results.

Recent advancements in EEG measurement technology have brought attention to the analysis of high-frequency ranges. These technological improvements have made it possible to analyze signals in the high-frequency band, which was previously challenging, and provide an opportunity to explore distinctive dynamic patterns in this range with greater precision. However, the high-frequency band is also prone to strong interference from external signals, such as electromyography (EMG) artifacts. These artifacts can reduce the reliability of the analysis. Therefore, to effectively extend and apply the approach proposed in this study, further research on artifact removal techniques in the high-frequency band is required.

Conventional frequency-based analysis methods, such as Fourier Transform (FFT) and Wavelet Transform (WT), each have their own advantages. However, they also have certain limitations when analyzing nonstationary signals.

Limitations of FFT are as follows:

Loss of Time Information: FFT transforms a signal into its frequency components but fails to capture temporal variations. This limitation makes it difficult to analyze transient frequency changes in nonstationary signals.

Restricted Frequency Resolution: The frequency resolution of FFT depends on the entire signal length, making it challenging to accurately analyze frequency variations occurring within short time windows.

Limitations of Wavelet are as follows:

Difficulty in Scale Selection: While Wavelet Transform allows signal analysis at multiple scales, the choice of an appropriate scale significantly affects the results. Poor scale selection may lead to the loss of important information.

High Computational Complexity: Compared to FFT, Wavelet Transform requires more computational resources, which can be a drawback for real-time analysis or large-scale data processing.

Despite these limitations, FFT and Wavelet Transform remain widely used because they offer intuitive interpretation and easy visualization. In other words, even though more advanced analytical methods exist, the ability to easily understand and utilize visually interpretable results is a key advantage, allowing these conventional methods to continue being widely applied.

DMD was utilized to effectively analyze the high-frequency range of nonstationary signals, overcoming the limitations of traditional methods. Since DMD provides both time and frequency information simultaneously, it is well-suited for capturing the dynamic variations of signals. It enables the extraction of self-similar high-frequency dynamics in nonstationary signals, offering more reliable results than conventional frequency-based analysis methods. Additionally, DMD retains the visual interpretability of traditional approaches while addressing the limitations of FFT and Wavelet Transform. In this study, even when analyzing only half of the high-frequency range instead of the full range using DMD modes, high classification accuracy was achieved. Further experiments confirmed that minor variations in the number of selected features ($v$ in (13)) did not significantly affect the results. Ultimately, DMD preserves the interpretability of conventional methods while enabling more effective analysis of neural dynamics in the high-frequency range.

In conclusion, significant dynamic changes observed in the high-frequency range above the gamma wave in nonstationary EEG signals can be effectively reflected as distinct patterns in the DMD modes. This approach presents a new methodology that can complement the limitations of traditional low-frequency-based EEG analysis techniques. Exploring the potential applications of these unusual patterns observed in the high-frequency range, such as in the diagnosis of neurological disorders or the evaluation of treatment responses, will be an important area of future research.

In this study, the train/test dataset officially provided by Kaggle was used without additional cross-validation. Since this dataset has already been utilized and validated in numerous previous studies, we consider its reliability to be well established. The dataset was provided in a predefined split, and no modifications or alterations were made to the data partitioning during the research process. Therefore, the possibility of data leakage was controlled by Kaggle's predefined data partitioning, ensuring the reproducibility and objectivity of the experimental results. It should be noted, however, that this partition divides trials within participants rather than across them, so recordings from the same individual appear in both the training and test sets. Subject-wise schemes such as leave-one-subject-out cross-validation have been shown to yield lower and more realistic estimates of cross-subject generalization in EEG disorder classification, since epoch-level splits allow a classifier to exploit subject identity rather than disease-related structure [16, 31]. The accuracy reported here should therefore be interpreted as an upper bound obtained under within-subject evaluation. This is particularly important because the primary objective of this study is not merely to achieve high classification accuracy but to verify the existence of distinguishable neural dynamics in the high-frequency range. However, to objectively evaluate the model's performance, future studies should consider using data from different participants as the test set, applying cross-validation, and conducting comparative experiments with various classifiers.

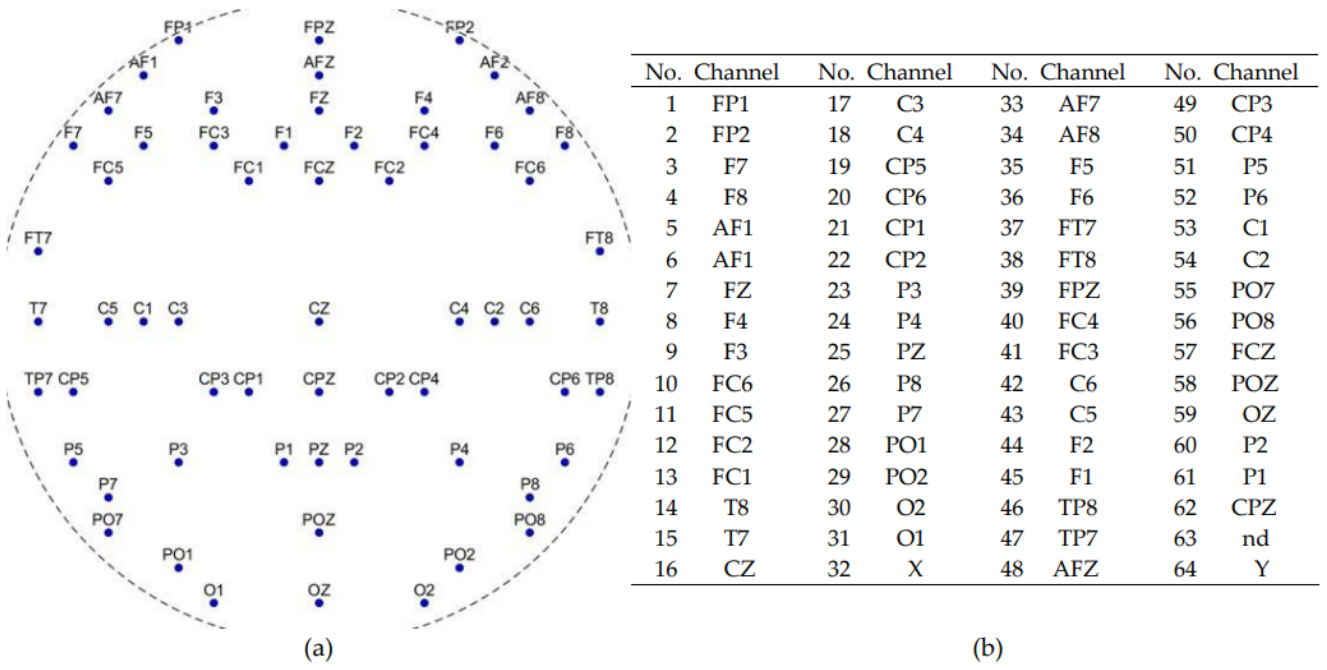

| No. | Channel | No. | Channel | No. | Channel | No. | Channel |
|---|---|---|---|---|---|---|---|
| 1 | FP1 | 17 | C3 | 33 | AF7 | 49 | CP3 |
| 2 | FP2 | 18 | C4 | 34 | AF8 | 50 | CP4 |
| 3 | F7 | 19 | CP5 | 35 | F5 | 51 | P5 |
| 4 | F8 | 20 | CP6 | 36 | F6 | 52 | P6 |
| 5 | AF1 | 21 | CP1 | 37 | FT7 | 53 | C1 |
| 6 | AF1 | 22 | CP2 | 38 | FT8 | 54 | C2 |
| 7 | FZ | 23 | P3 | 39 | FPZ | 55 | PO7 |
| 8 | F4 | 24 | P4 | 40 | FC4 | 56 | PO8 |
| 9 | F3 | 25 | PZ | 41 | FC3 | 57 | FCZ |
| 10 | FC6 | 26 | P8 | 42 | C6 | 58 | POZ |
| 11 | FC5 | 27 | P7 | 43 | C5 | 59 | OZ |
| 12 | FC2 | 28 | PO1 | 44 | F2 | 60 | P2 |
| 13 | FC1 | 29 | PO2 | 45 | F1 | 61 | P1 |
| 14 | T8 | 30 | O2 | 46 | TP8 | 62 | CPZ |
| 15 | T7 | 31 | O1 | 47 | TP7 | 63 | nd |
| 16 | CZ | 32 | X | 48 | AFZ | 64 | Y |

(b)

**Figure 6.** (a) The locations of the EEG electrodes on the cortex. The label and coordinate for each electrode were referenced from Table 1 in [33]. (b) List of Channels.